\documentclass[lettersize,journal]{IEEEtran}
\usepackage{amsfonts}
\usepackage{times}
\usepackage{graphicx}
\usepackage{latexsym}
\usepackage{dsfont}
\usepackage{amssymb}
\usepackage{amsmath}
\usepackage{cite}
\usepackage{verbatim}
\usepackage{subfigure}

\newcommand{\figref}[1]{{Fig.}~\ref{#1}}

\def\bb0{{\mathbb{0}}}

\def\bb{{\mathbf{b}}}

\def\b0{{\mathbf{0}}}

\def\cB{\mathcal{B}}

\def\cE{\mathcal{E}}

\def\sf0{{\mathsf{0}}}

\usepackage{epstopdf}
\usepackage{enumerate}
\usepackage{algorithm}
\usepackage{amsmath}
\usepackage{float}
\usepackage{hyperref}
\usepackage{color}
\usepackage{makeidx}
\usepackage{bm}
\usepackage{cleveref}
\usepackage{url}
\usepackage{steinmetz}
\usepackage{multirow}
\usepackage{varwidth}
\usepackage{soul}
\usepackage{xcolor,colortbl}
\usepackage{tabularray}

\begin{document}

\title{Integrated Sensing and Communication for 6G: \\ Ten Key Machine Learning Roles}

\author{Umut Demirhan and Ahmed Alkhateeb 
\thanks{The authors are with the School of Electrical, Computer and Energy Engineering, Arizona State University, (Email: udemirhan, alkhateeb@asu.edu).}}

\maketitle

\begin{abstract}
Integrating sensing and communication is a defining theme for future wireless systems. This is motivated by the promising performance gains, especially as they assist each other, and by the better utilization of the wireless and hardware resources. Realizing these gains in practice, however, is subject to several challenges where leveraging machine learning can provide a potential solution. This article focuses on ten key machine learning roles for joint sensing and communication, sensing-aided communication, and communication-aided sensing systems, explains why and how machine learning can be utilized, and highlights important directions for future research. The article also presents real-world results for some of these machine learning roles based on the large-scale real-world dataset DeepSense 6G, which could be adopted in investigating a wide range of integrated sensing and communication problems. 
\end{abstract}

\section{Introduction}

Integrating sensing and communication is envisioned as a key pillar for 6G systems and is attracting increasing interest at both academia and industry \cite{liu2021integrated,Zhang2021,Liu2022,kumari2017ieee}. This is thanks to the potential gains this integration is expected to offer for the better utilization of the spectrum and the reduction in the hardware cost and power consumption. Further, if properly designed, the sensing and communication functions can benefit each other. For example, the sensory data of the environment can be leveraged to aid the communication beam management and resource allocation \cite{demirhan2022beam}, and the communication of the sensing data can enable real-time network sensing \cite{liu2021integrated}. Achieving this vision of integrated sensing and communication (ISAC) systems, however, requires overcoming several challenges. For example, this integration complicates the design of the communication/signal waveform, the allocation of the system and hardware resources, the management of the interference, and the overall network operation. All that motivates the research for novel approaches that overcome these challenges and enable the ISAC system gains in real-world deployments.

Machine learning has powerful capabilities in extracting and leveraging temporal/spatial patterns, approximating complex models, and solving difficult optimization problems. In the last few years, these interesting capabilities found many important applications in wireless communications and sensing. For example, deep learning models have been utilized to optimize the modulation and coding schemes in an end-to-end fashion \cite{felix2018ofdm}, learn site-specific beam codebooks for millimeter wave systems \cite{zhang2022reinforcement}, and optimize the wireless resource allocation in complex network settings \cite{Sun2018}. For wireless sensing systems, machine learning has already been used in enhancing the sensing/imaging resolution and improving the pattern detection and classification quality, which is unlocking novel applications in security, healthcare, etc. \cite{zhang2020device}. It becomes natural then to ask about the potential of leveraging machine learning to overcome the ISAC system challenges. 

In this article, we present ten key roles for machine learning in ISAC systems, organized in the three main directions of joint sensing and communication (JSC), sensing-aided communication, and communication-aided sensing. 

For \textbf{joint sensing and communications}, where the communication and sensing functions share the same hardware and/or wireless resources, machine learning has the potential of (i) optimizing the waveform design, (ii) learning site-specific spatial sensing and communication beam patterns, (iii) enabling self-interference cancellation at the JSC transceivers, (iv)  optimizing the wireless and system resource allocation,  (v) enhancing the JSC system security, and (vi) enabling the network-level operation of the JSC systems.

In the context of \textbf{sensing-aided communication}, where the sensory data could be leveraged to enhance the wireless communication operation, machine learning can find important roles in (vii) facilitating sensing-aided beam prediction/tracking in large antenna array systems and (viii) enabling sensing-aided blockage prediction and proactive hand-off in millimeter wave and sub-terahertz systems, where line-of-sight (LOS) link blockages represent critical challenges for the reliability and latency of the network.

Finally, for the \textbf{communication-aided sensing} direction, where the communication signals and systems are utilized to extend the sensing capabilities, machine learning can have important gains in (ix) optimizing the communication-aided sensing functions and in (x) enabling communication-aided network/distributed sensing.

The goal of this article is to expose the potential of leveraging machine learning in various ISAC problems. In the next sections, we discuss the ten key roles and highlight important directions for future research. Further, for some of these roles, we demonstrate the performance gains using large-scale datasets comprising co-existing sensing and communication data that are  based on real-world measurements.

\begin{table*}[]
	\label{tab:roles}
	\caption{ISAC Machine Learning Roles and Examples}
	\begin{center}
		\SetTblrInner{hlines, vlines, vspan=\line}
		\begin{tblr}{Q[c,m,0.1\linewidth]Q[c,m,0.016\linewidth]Q[c,m,0.15\linewidth]Q[c,m,0.635\linewidth]}
			\hline
			\textbf{Category} & \SetCell[c=2]{c}\textbf{Role} & & \textbf{Example} \\ 
			\hline
			\SetCell[r=6]{c} {Joint Sensing\\and\\Communication} & 1 & {Waveform\\optimization} & Active learning could be leveraged to adaptively optimize the JSC waveforms for maximizing the target detection probability and the achievable communication data rates. \\ 
			\cline{2-4} 
			& 2  & {Learning spatial\\beam patterns} & JSC systems can utilize online reinforcement learning models to self-configure their beam patterns to match the  environment and hardware imperfections using power measurements as rewards.  \\ 
			\cline{2-4} 
			& 3 & {Self-interference cancellation\\for full-duplex} & {Non-linear effects that are difficult to accurately model could potentially be learned with supervised DNN models. For instance, a DNN that takes the transmit signal as input could learn how to approximate the non-linear transmit/receive  and channel models to predict the self-interference.} \\ 
			\cline{2-4} 
			& 4  & {Resource\\optimization} & Leveraging reinforcement learning, JSC system observations such as the achievable data rate and target detection accuracy can be used to dynamically allocate and optimize the sensing and communication power, time, frequency, and beam resources. \\
			\cline{2-4} 
			& 5   & {Enhancing\\system security} & High level parameters, such as the locations of the detected objects and data rates, can be monitored to detect the attacks (anomalies) on both sensing and communication by using autoencoders. \\ 
			\cline{2-4} 
			& 6  & {Enabling\\network operation} & Multi-agent reinforcement learning can be utilized to jointly optimize the JSC beamforming vectors and power allocation with minimal coordination. \\ 
			\hline
			\SetCell[r=2]{c} {Sensing\\Aided\\Communication}    & 7 & {Sensing aided\\beam prediction} & Recurrent neural networks utilizing the sensing measurements, such as range, angle, and Doppler, over time can be used to predict future beams in highly-mobile scenarios. \\ 
			\cline{2-4} 
			& 8 & {Sensing aided blockage\\and hand-off prediction} & Attention neural networks can be utilized to predict the future locations of the users and possible blocking objects, and hence proactively predict LOS link blockages. \\ 
			\hline
			\SetCell[r=2]{c} {Communication\\Aided\\Sensing}    & 9  & {Communication-aided\\sensing\\optimization} & By adopting image-like representations of the sensing and communication measurements, super-resolution and denoising CNN models can be leveraged. For example, low-resolution noisy spatio-temporal channel maps can be fed to a CNN to obtain high-quality maps. \\ 
			\cline{2-4} 
			& 10 & Communication-aided network sensing &  Graph neural networks present a scalable architecture for combining the radar measurements from different network nodes to achieve the network sensing over communication functions. \\ 
			\hline
		\end{tblr}
	\end{center}
\end{table*}

\section{Joint Sensing and Communication}

The communications and radar sensing functions can be jointly applied by fully or partially sharing the same hardware, waveform, and wireless resources. This, however, introduces several challenges where machine learning could potentially play key roles. Next, we present some of these machine learning roles in JSC systems.

\begin{figure*}[t]
	\centering
	\includegraphics[width=1\linewidth]{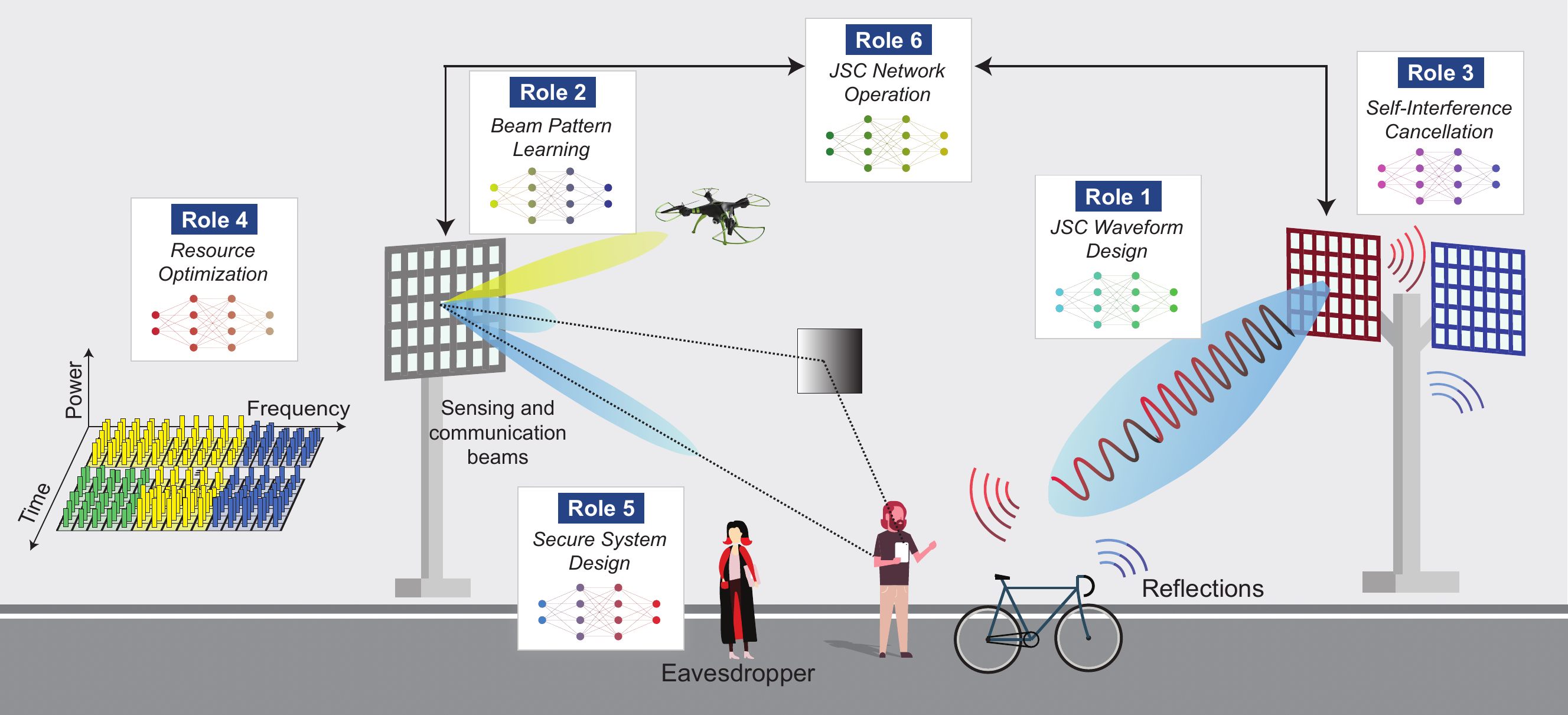}
	\caption{Illustration of the machine learning roles in JSC systems. Using the observations from the JSC devices (e.g., channel measurements), the machine learning models can aid the design tasks and enable novel capabilities.}
	\label{fig:JSC}
\end{figure*}

\textbf{Role 1: JSC waveform design:}  Using a single waveform for both the radar and communication functions is a main goal for JSC systems \cite{liu2021integrated,kumari2017ieee}. In this case, the joint waveform should be carefully designed to  meet the communication objectives (e.g., data rates, latency, and reliability) while enabling the desired sensing capabilities. Efficiently achieving that, however, is a challenging research problem because of the trade-offs between the sensing and communication operations. This becomes particularly challenging when (i) accounting for the various communication and sensing system requirements, such as the peak-to-average power ratio constraints of the OFDM systems and the  signal-to-clutter-plus-noise ratio for the sensing, (ii) considering the hardware limitations, such as the ADC/DAC quantization and the power amplifier non-linearity, and (iii) extending to highly-mobile scenarios, where the radar and communication channels are rapidly varying.

Machine learning has the potential to optimize the JSC waveform design through both model and data driven approaches. One particular promising direction is to design hybrid conventional signal processing and machine learning solutions that learn how to solve complex JSC waveform design problems. These solutions could leverage the domain knowledge about radar and communication systems to extract useful features for the \textit{multi-objective} machine learning models, hence reducing  their complexity and lowering their inference latency. Self-learning \textit{site-specific} JSC waveforms is another promising gain of machine learning; leveraging machine/deep learning could enable the online optimization of the waveforms to adapt to the deployment scenario and adopted hardware \cite{zhang2022reinforcement}. This has the potential of enhancing the overall JSC system performance compared to generic waveforms that do not account for the dependency on the environment geometry, user location, mobility pattern, and hardware imperfections. Another interesting approach for the same objective is the use of auto-encoder models for end-to-end waveform learning \cite{felix2018ofdm}, which can be useful when the hardware models are unknown. 


\textbf{Role 2: Learning spatial beam patterns:}  Deploying multiple antennas at the transmitters and/or receivers is a key characteristic of current and future wireless communication and sensing systems. To realize the gains of these multiple-antenna systems, however, it is essential to efficiently design their beamforming vectors. For JSC systems, the beam design criteria are different for the communication and sensing functions. For example, the communication systems could prioritize designing the beams to focus the transmitted signal power towards the receiver.  Meanwhile, the radar operation may prefer having a wider beamwidth to capture more sensing information about the various scatterers in the environment. Therefore, designing unified JSC beam patterns requires compromising between the sensing/communication functions. This becomes particularly challenging with the strict hardware constraints, such as those imposed on analog/hybrid architectures.

There are several ways where machine learning can aid the JSC beamforming design problem.  First, machine/deep learning could help reduce the complexity of the JSC beam design problems by learning the structure of the communication/sensing MIMO channels. This could lead to dimensionality reduction, which is particularly useful for large-scale MIMO systems. In addition,  online learning models can self-learn site-specific spatial beam patterns and codebooks that are optimized specifically for the site geometry, hardware, and deployment scenario. This could potentially realize superior sensing and communication capabilities, especially for hardware-constrained  MIMO systems, such as analog/hybrid  architectures. While such gains have been demonstrated before for communication systems  \cite{zhang2022reinforcement}, their potential advantages for JSC beams is still to be investigated.

\textbf{Role 3: Self-interference cancellation:} Full-duplex operation is a key enabling technology for joint communication and sensing systems \cite{barneto2021full}. Specifically, the JSC functions may require the transmitter and receiver to operate concurrently in a full-duplex mode. For example, the transmitter could be transmitting the JSC waveform while the receiver is capturing the backscattered/reflected signals for the radar processing. This requires, however, handling the self-interference at the JSC transceiver which is a major challenge. Canceling the self-interference at the antenna, analog, or digital domains demand sufficient knowledge about the circuit non-linear models, antenna coupling, channels, which is typically hard to acquire. Further, the implementation of these self-interference cancellation techniques (circuits and signal processing) is generally associated with high complexity and strict constraints. 

Machine learning, and in particular DNNs, could potentially help in building full-duplex JSC systems. For example, the DNNs can approximate the complex self-interference models. This is motivated by  their powerful capabilities of learning non-linear functions from the available data, without requiring accurate knowledge of the underlying models. At the antenna domain, machine learning could be utilized to learn transmit/receive beamforming patterns for self-interference nulling, while accounting for the coupling and hardware imperfections. Machine learning models can also be designed to reduce the self-interference cancellation complexity, and hence enable real-time processing, especially when implemented using specialized computing hardware. 

\textbf{Role 4: Resource optimization:} JSC systems can, in general, either adopt a single waveform for simultaneous communication and sensing over the same time/frequency resources or separate waveforms. In the latter case, it becomes important to efficiently allocate the various wireless resources, such as time, frequency, and power, between the sensing and communication functions. This is challenging because of the different trade-offs between the sensing quality and the communication objectives. Additional challenges on the service scheduling are imposed when the JSC systems use massive MIMO systems. For example, with directional transmission/reception, the time-frequency resources are limited to the communication users or radar targets in the transmission/reception directions. 

Machine learning has the potential to enable better utilization of the resources by intelligently allocating them between sensing and communication tasks. By leveraging  prior observations and  side information about the context and user activity,  learning models can make more efficient resource optimization decisions. For example, the distribution and information about previous user requests could be useful in inferring the future ones, and hence proactively allocating their resources.  Reinforcement learning approaches could also efficiently learn these distributions while navigating through the high-dimensional resource optimization space for high-performing solutions. This can also reduce the dataset collection requirements by relying on highly-quantized rewards. 

 
 \begin{figure*}[!t]
 	\centering
 	\includegraphics[width=1\linewidth]{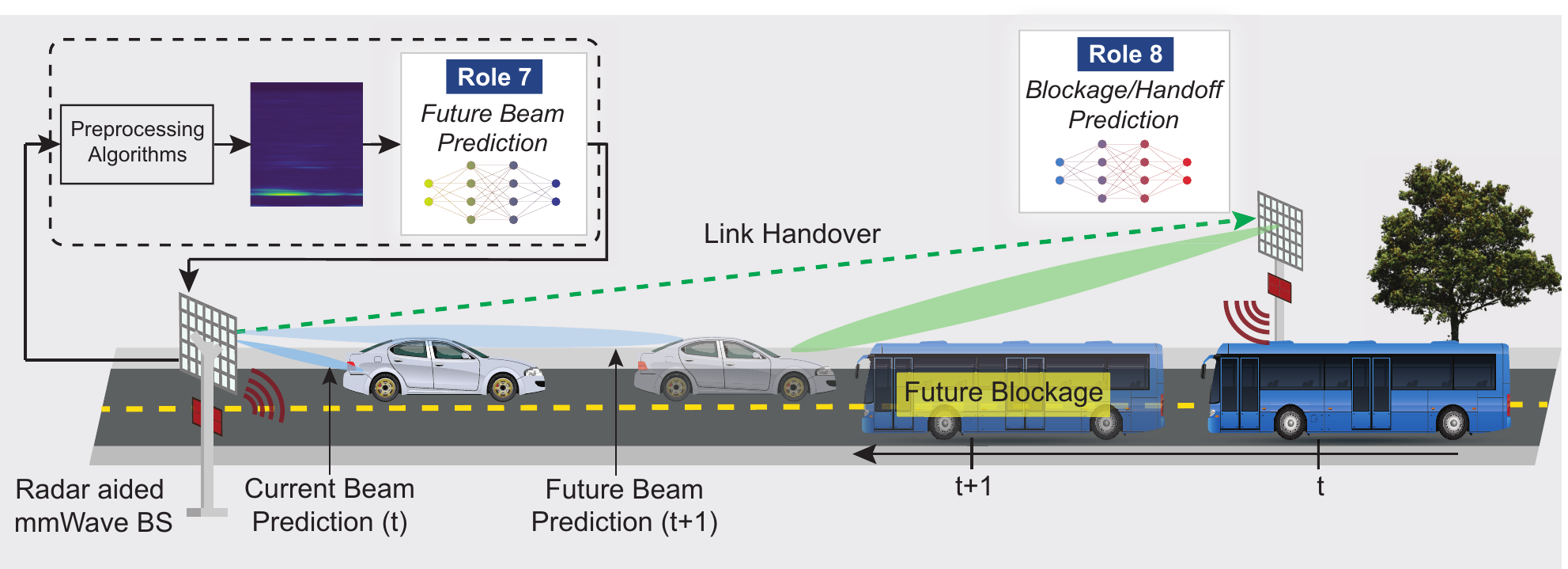}
 	\caption{Sensing aided beam and blockage prediction with machine learning. The pre-processed radar signals can be fed to machine learning models for predicting current/future beams and LOS link blockages. }
 	\label{fig:sensingaidedbeam}
 \end{figure*}

\textbf{Role 5: Enhancing JSC system security:} In the presence of malicious agents, such as an eavesdropper or attacker, the communication systems need be designed while accounting for certain security objectives \cite{liu2021integrated,Zhang2021}. For example, the JSC waveforms and the transmit beamforming vectors may need to be designed in a way that makes it hard for the eavesdroppers to identify the transmitted signals and obtain meaningful messages. This normally results in complex waveform optimization problems. Further, realizing secure JSC systems may require these systems to be able to detect the presence of attackers  and identify their types/roles. This is also a difficult objective for the JSC systems since they  need to do that while maintaining the sensing and communication functions. In addition, various types of attacks may require different and adaptable detection mechanisms. 

Machine learning can enhance the security of the JSC systems with their inherent capability of extracting and utilizing information from different domains. For example, machine learning models can more accurately and robustly extract the channel information of the attackers from the sensing data, and efficiently utilize it to enhance the security of the JSC system operation. Machine learning models also have powerful capabilities in detecting anomalous behavior. This can be leveraged, for instance, to detect and predict the presence of malicious attacks. From system complexity perspective, machine learning could help in relaxing the increased complexity of the JSC system design under security constraints. For example, online beam pattern learning approaches could be leveraged to optimize the beam patterns and shape nulls towards the attackers direction using only power measurements and without requiring explicit knowledge about their locations, channels, or hardware models.

\textbf{Role 6: Enabling JSC network operation:} 
The JSC nodes will eventually need to operate together in a network setting \cite{liu2021integrated,Zhang2021,Liu2022}. Considering multiple JSC nodes simultaneously operating in the same frequency band, however, brings several important challenges. On one hand, the various devices may cause considerable interference that affect the performance of both the communication and sensing functions. For that, the JSC waveforms, resource optimization, beamforming, and sensing/communications post-processing need to be optimized to manage the interference between devices, which normally yields complex problems. On the other hand, if there is coordination between the JSC nodes, this may help improve the sensing and communication performance. For example, the sensing receivers could take advantage of the transmissions from the other devices in the network via multi-static radar processing and the communication devices could leverage the coordination for better coverage and data-rate. Realizing these interesting gains, in practice, however requires sufficient knowledge about the channels between the different JSC nodes and solving complex joint optimization problems.

Machine learning can allow efficient processing of network level transmissions, managing coordination with a low complexity, and presenting interference-management solutions.  For example, distributed reinforcement learning based solutions, with multiple agents deployed at the distributed nodes, has the potential to enable network level management with minimal coordination. Such decentralized reinforcement learning solutions can also adapt well to the dynamic environments thanks to their exploration capabilities. Further, distributed machine learning approaches can leverage the spatial structure of the wireless communication and sensing channels in a particular site/deployment to develop optimized network-level waveforms and  signal processing functions. To this end, it is interesting to take advantage of  advanced distributed/privacy preserving learning methods, such as federated learning \cite{kairouz2021advances}, in designing joint sensing and communication systems.

\begin{figure*}[!t]
	\centering
	\includegraphics[width=.9\linewidth]{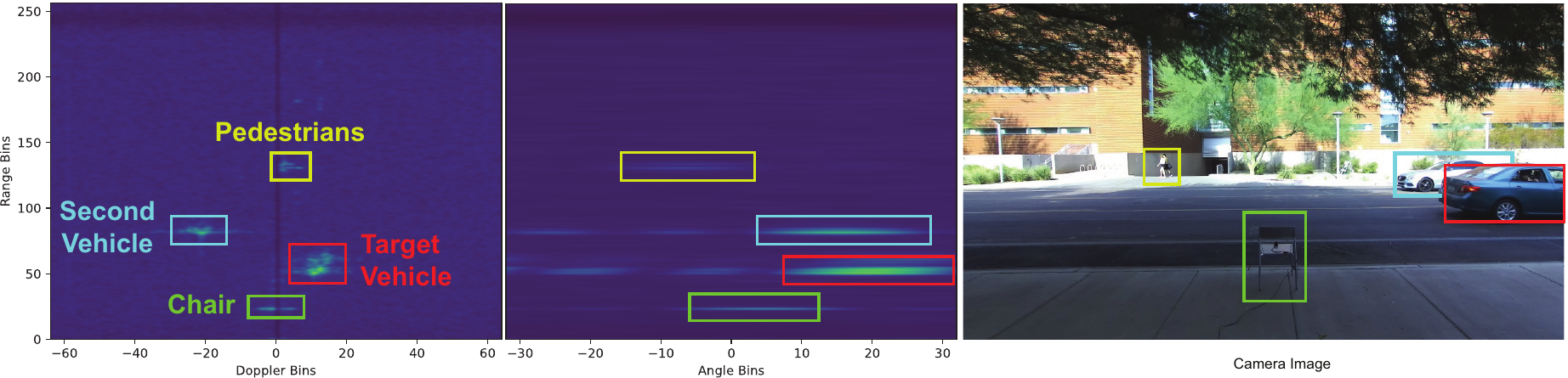}
	\caption{The radar range-angle and range-velocity maps of a data sample from Scenario 9 of the DeepSense 6G dataset \cite{deepsense} are shown along with the camera image. Different elements in the environment, including pedestrians and two vehicles,  are identified in the images.}
	\label{fig:beampredictionsample}
\end{figure*}

\begin{figure}[!t]
	\centering
	\includegraphics[width=1\linewidth]{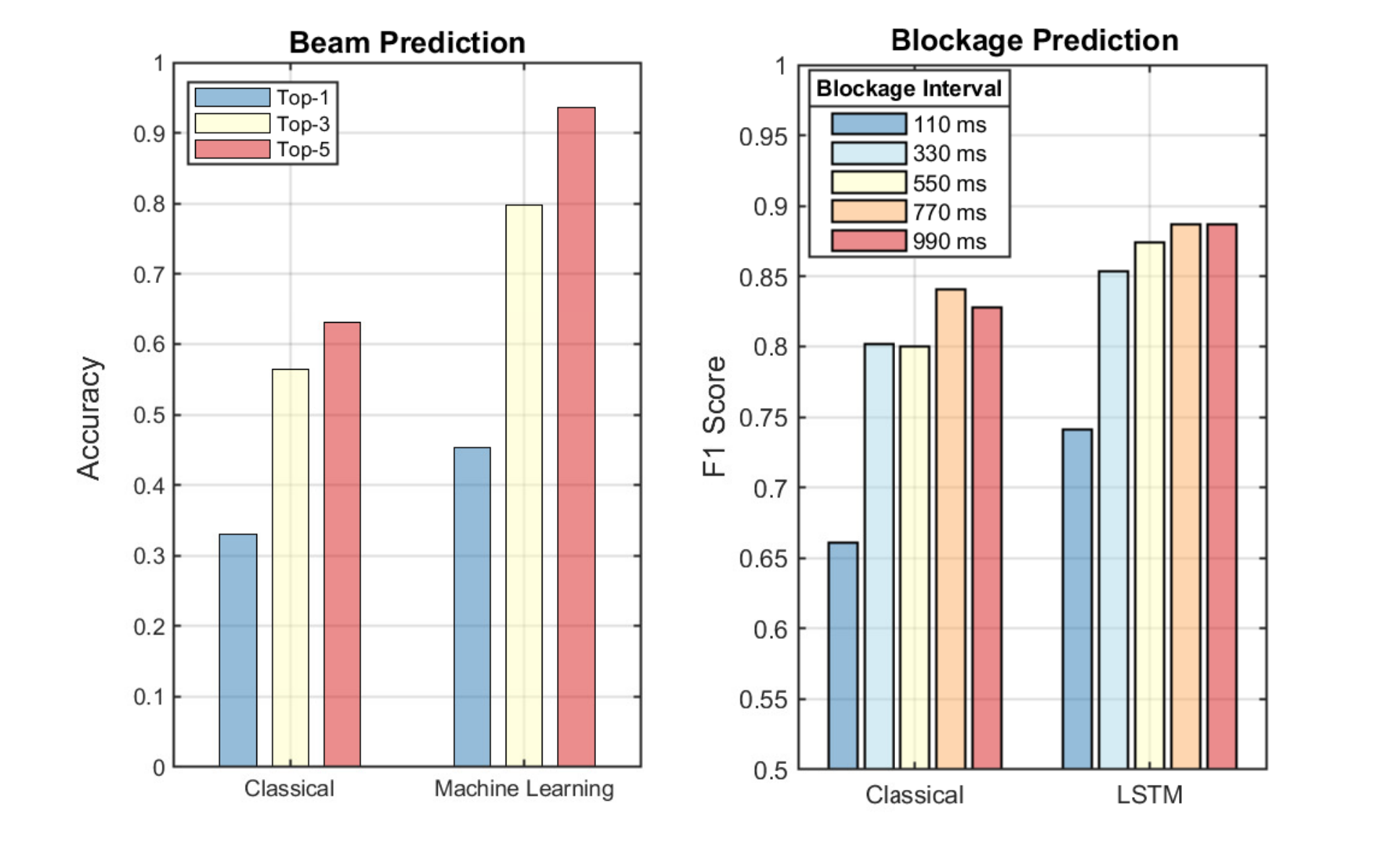}
	\caption{Real-world sensing aided beam and blockage prediction performance with the FMCW radar aided communication solutions in \cite{demirhan2022beam, demirhan2022blockage}.}
	\label{fig:beamblockageresults}
\end{figure}

We would like to note here that one of the main challenges with all of the ISAC roles is defining the right performance metric. For example, in JSC, combined with a communication target, various types of radar objectives, such as information-theoretic, detection, estimation or annotation based metrics, have been utilized in the literature. When adopting machine learning to address these ISAC challenges, it is important to develop loss functions that accurately captures the performance metrics of interest. Another important consideration for all these machine learning applications in ISAC is the availability of datasets. Recent dataset releases, such as DeepSense 6G \cite{deepsense}, which will be discussed shortly, could be an initial step towards this objective. Further, in practical ISAC systems, the availability of annotations could play an important factor is selecting the machine learning approach. For instance, unsupervised DNNs can be adopted for complex problems with unknown labels, supervised DNNs for problems with known optimal solutions/annotations, and semi-supervised learning when a small number of annotations is available. 

\section{Sensing Aided Communication}

The  radar sensing could provide useful awareness about the communication environments, which could potentially be utilized by the communication systems to improve their performance. For example, a mmWave basestation could use the awareness about the location and mobility pattern of the mobile user/scatterers to guide the beam alignment or proactively predict LOS link blockages.  This motivates adding radar sensors to the communication infrastructure nodes and UEs or leveraging the sensors that may already exist. In this \textit{sensing aided communications} application, the radar and communication systems may not necessarily need to operate in the same band or utilize the same hardware. For example, a 28GHz communication basestation can utilize the  sensing  information acquired by a frequency modulated continuous wave (FMCW) radar device operating at the automotive 77GHz band. In this section, we explore the potential machine learning roles in this sensing-aided communication direction.

\textbf{Role 7: Sensing aided beam prediction:} 
Moving to the higher frequency bands (mmWave and sub-terahertz) is a key characteristic of current and future wireless communication systems due to the larger available bandwidth at these frequency bands. To maintain sufficient receive signal power, these systems need to deploy large antenna arrays and use narrow beams. Aligning the narrow beams at the transmitters and receivers, however, is associated with high beam training overhead that scales with the number of antennas. Interestingly, this beam alignment highly depends on the locations of the transmitters/receiver and the geometry around them. This motivates leveraging radar sensors to provide useful information about the locations/geometry of the communication environment, and hence, guide the beam selection decisions and eliminate or significantly reduce the beam training overhead \cite{Ali2020,liu2021integrated,demirhan2022beam}. Achieving that in practice, however, is challenging as it requires calibrating the radar and communication systems, addressing clutter and background noise, and differentiating between the target user and the other distracting objects.

Leveraging the recent advances in machine/deep learning provides a promising path for overcoming the sensing-aided beam prediction challenges \cite{demirhan2022beam}. First, the powerful capabilities of deep neural networks in learning and approximating complex models could be pivotal in efficiently learning the radar sensing data-to-beam mapping. This mapping could, in practice, be complex and highly non-linear due to (i) the various mismatches between the sensing and communication systems (e.g., hardware and fields of view), and (ii) the different noise sources in the sensing and beam measurements. In addition, the advances in transfer-, meta-, and federated-learning could also be harnessed to enable a fast and scalable deployment, and to reduce the calibration and dataset collection overhead. Leveraging the domain knowledge in wireless communication and sensing is critical in the design of training methodologies. For example, transferring the learning of a sensing-aided beam prediction model may benefit from pre-processing the input radar data to account for the device orientation changes.  For further reduction in the beam training overhead, advanced object tracking and time-series prediction solutions could also be utilized for future beam prediction (also called beam tracking). For example, a sequence of radar samples could be used to predict the transmit/receive beams for the next (future) few time samples by using the recurrent neural networks.

\textbf{Real-World Evaluation:} A demonstration of the radar aided blockage prediction application using a large-scale real-world dataset was first presented in \cite{demirhan2022beam}. Specifically, in this work, a real-world dataset for a vehicle-to-infrastructure scenario was constructed (as part of the the DeepSense 6G dataset \cite{deepsense}). In this scenario, a  vehicle carrying an omni-directional mmWave transmitter passes by the receiver unit, which is stationary on the side of the road. This receiver unit comprises a $60$ GHz mmWave antenna array with a $64$ beam codebook, and a $77$ GHz FMCW radar. The dataset consists of nearly six thousand samples, where each sample contains the power measurements of each beam and a raw radar measurement. A sample from the dataset is presented in \figref{fig:beampredictionsample}.

For this dataset, two radar-aided beam selection solutions have been developed in \cite{demirhan2022beam}: (i) A classical solution that relies on a detection algorithm based on the range-angle maps extracted from the raw radar measurements, and a look-up table that maps the detected vehicle to the beams. (ii) A deep learning solution that takes the range-angle maps as the input and returns the most likely beams. In the left side of \figref{fig:beamblockageresults}, we present the accuracy of the results. In particular, the top-$k$ beam prediction results are shown on the left figure. The machine learning solution provide $\sim44\%$ top-$1$ accuracy outperforming the classical solution by $\sim10\%$. Moreover, the top-$5$ accuracy of the machine learning reaches up to $93\%$, which shows a significant potential for real-world applications. This, together with the reasonable latency \cite{demirhan2022beam}, highlights the potential of leveraging machine learning for real-world sensing aided beam prediction applications.

\textbf{Role 8: Sensing aided blockage and hand-off prediction:} 
The signal propagation at the high frequency bands is subject to high penetration loss and diffuse scattering. Therefore, the mmWave and sub-terahertz communication networks rely mainly on LOS links and are very sensitive to blockages. This, however, highly challenges the reliability of these networks as almost any LOS obstructing object can yield high receive power degradation and sudden link disconnection. Further, handling link failures through \textit{reactive} hand-off or beam switching  is normally associated with high latency overhead. Traditional solutions for link blockages focused mainly on multiple connectivity, for example, by keeping the mobile user connected all the time to multiple basestations, even if only one link is needed. These multi-connectivity solutions, however, underutilizes the network resources and degrade the overall network performance.

Integrating radar sensing and machine learning could enable the \textit{proactive} prediction of the link blockages  \cite{demirhan2022blockage}. In particular, radar sensing provides valuable awareness about the communication environment, including the position, shape, and mobility features of the mobile user and the other static/dynamic objects in the environment. Leveraging this environment awareness, machine/deep learning models could potentially learn to predict whether or not a future LOS link blockage will happen in a proactive way, i.e., before it actually happens. Specifically, the objects that will cause the blockages and the time and duration of these blockages can be predicted using the sensing data with the help of the machine/deep learning models. This proactive blockage prediction could enable the mobile network to make proactive decisions on hand-off and beam switching to avoid the sudden disconnection of the communication session, leading to better reliability and lower latency.

\textbf{Real-World Evaluation:}  To show the potential of the sensing aided blockage prediction with machine learning in real-world application, we collected the scenario 30 of the DeepSense 6G dataset \cite{deepsense} in \cite{demirhan2022blockage}. The same setup in the previously described beam prediction is used with a static transmitter. During this data collection, several objects, e.g., vehicles, cyclists, pedestrians, were actively using the road and causing blockages to the communication transmission. With this real-world dataset, we developed two proactive blockage prediction solutions: (i) A classical approach where all the moving objects are tracked with a Kalman filter using the radar samples. The Kalman filter states, the positions and speeds of the objects, are utilized to predict future blockages with a k-nearest neighbors method. (ii) An LSTM based solution where a sequence of the radar observations are fed to a combination of CNN and LSTM to directly predict the future blockages. In the right side of \figref{fig:beamblockageresults}, we show the F1-score of the solutions. As seen in the figure, the LSTM solution is able to predict future blockages $330$, $550$, $770$, and $990$ms before they happen with F1-score that is more than $85\%$. These results show the potential of radar sensing and machine learning based future blockage prediction, which can enable highly-reliable and low-latency wireless  networks.

\section{Communication Aided Sensing}
The existing communications signals, such as the transmitted preamble, may also be utilized for sensing, which motivates the communications aided sensing concept \cite{taha2021millimeter,liu2021integrated,Zhang2021,kumari2017ieee}. In particular, the communication signals that are transmitted to a target receiver will also backscatter and reflect  from the various objects in the environment. By capturing these backscattered/reflected communication signals, one can opportunistically achieve sensing tasks such as positioning, object detection and tracking, and even imaging. It is important to note here that in the communications aided sensing framework, the sensing objective is not present in the design of the transmitted signals, and hence the (passive) sensing function is performed without penalizing the communication system performance or interfering with its operation, which is different than the JSC systems. This could be important for the scenarios where a higher priority should be given to the communication application or when we want to utilize the same communication hardware and frame structure for opportunistic sensing. Next, we discuss potential machine learning roles for communications aided sensing systems.

\textbf{Role 9:  Communication-aided sensing optimization:}  Processing the receive communication signals to realizing high sensing quality is challenging. This is because the communication signals and systems (frame structure, preamble, beams, hardware) are not optimized for the sensing function. For example, the impact of the clutter noise, multi-path reflections, and unoptimized side-lobes could distort the receive signals and challenge the design of the radar signal processing functions \cite{taha2021millimeter}. Further, classical signal processing approaches normally do not leverage prior observations (e.g., the information about the prior distribution of the propagation parameters) or make use of the side information (e.g.,  the knowledge about the static elements/scatterers in the environment). 

\begin{figure}[t]
	\centering
	\includegraphics[width=1\linewidth]{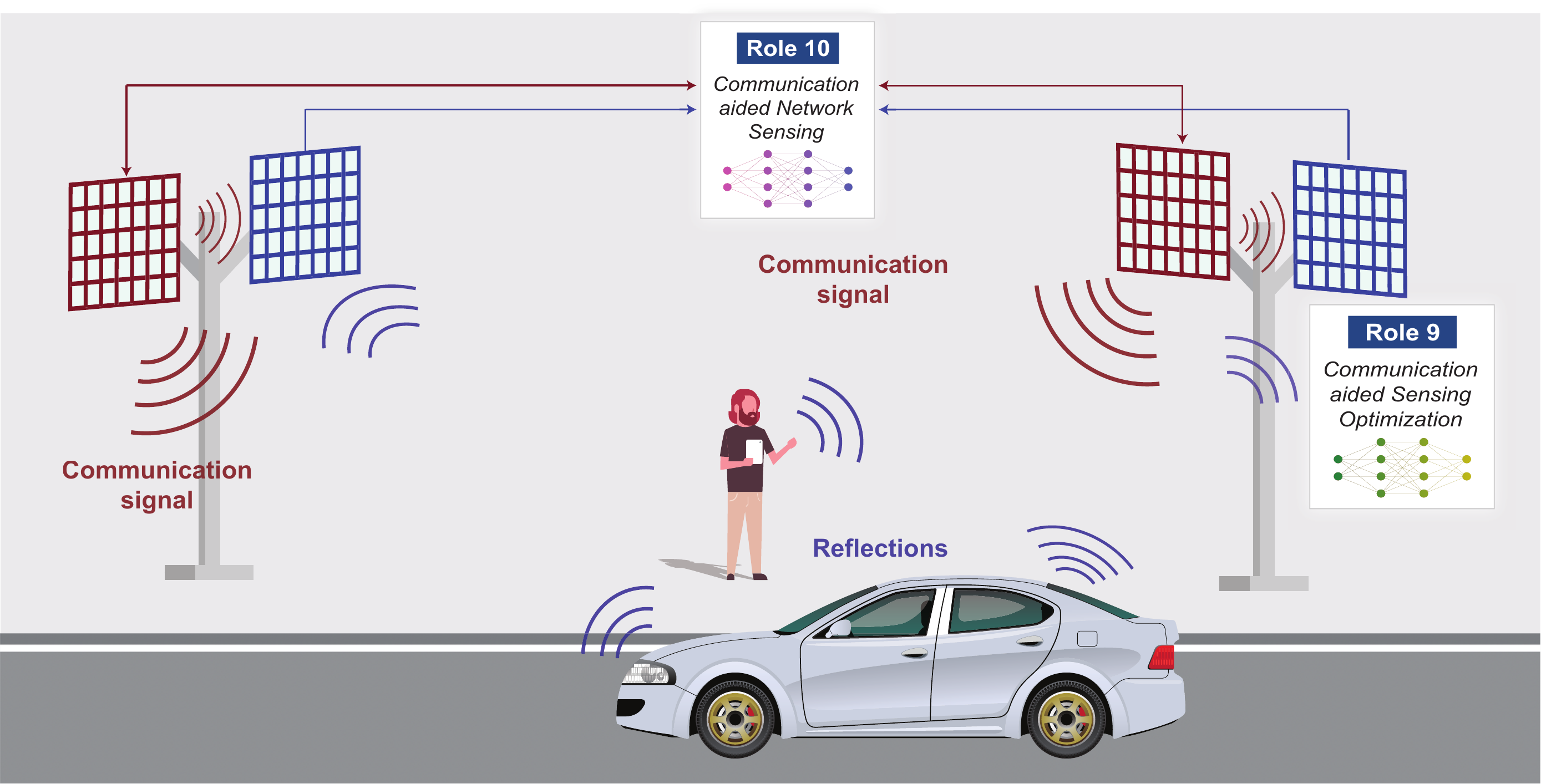}
	\caption{Illustration of the communication aided sensing tasks, where the communication transmissions are utilized with machine learning to improve the sensing quality. This can also be achieved at the network level, especially in  dense networks.}
	\label{fig:comm4sensing}
\end{figure}

Machine learning has the potential to enhance the communication aided sensing performance in multiple ways. First, by leveraging the powerful data-driven learning capabilities, the learning models could learn how to optimize the signal processing for the end sensing objectives without the need to explicitly model the various effects of the hardware/channels.  For example, the learning models may be able to approximate the complex underlying multipath signal propagation, which is typically hard to model, and leverage rich scattering to further extend the sensing range and resolution. Further, by adopting the sensing measurements along with the estimation of the communication channels, super-resolution and denoising CNN models can be leveraged. For example, high-quality range-angle maps can be constructed by these inputs. Similarly, the success of the machine learning applications in sparse recovery problems can be extended for the radar channel estimation given the observations. In addition, the side information about the communication environment could be implicitly captured in the training data and utilized by the  learning models to realize efficient \textit{environment-aware} sensing solutions.

\textbf{Role 10:  Communication-aided network sensing:}  The distribution of the communication equipment over an area, and the joint design and processing of the transmit/receive signals have interesting coverage and data rate gains. Such distributed communication systems could also be utilized for sensing. For example, the joint transmissions from multiple communication transmitters (multiple viewpoints) could be utilized for better sensing capabilities. Similarly, the communication signals captured at distributed receivers can be jointly processed for enhanced sensing performance. Designing  joint sensing solutions for distributed systems, however, is a challenging task, as these solutions need to account for the implications that result from the distribution of the transmission/sensing equipment, such as the synchronization requirements between these distributed nodes, the control overhead, and the complex design problems of the joint sensing processing functions.

The data-centric and model-free machine learning approaches can allow the distributed sensing tasks to utilize the communication signals and hardware without the need to accurately model these systems and their practical non-idealities. 
In addition, advanced distributed learning models and techniques \cite{kairouz2021advances}, such as transfer-learning, federated learning, and graph neural networks, can facilitate the use of distributed sensing devices and enable the efficient integration of the information gathered from different locations. These models and techniques could also help in compressing the distributed sensing data and reduce the coordination overhead. All that can enable the scalability of the distributed/network sensing functions while maintaining the possible complexity and data privacy constraints. 

\section{Conclusion}

Harvesting the gains of integrating sensing and communication requires overcoming the complexity of the system and algorithm design, efficiently allocating the wireless and hardware resources, managing the increased interference and control overhead among other challenges. In this article, we have attempted to highlight why and how machine learning could play important roles in addressing these challenges, which motivates future research to deeply investigate and fully realize these promising machine learning gains.


\end{document}